 \documentclass[prd,twocolumn,letterpaper,nofootinbib]{revtex4}
 \usepackage{graphicx}
 \usepackage{epstopdf}
 \usepackage{amsmath,amssymb}
\newcommand{\apjl}{Astrophys. J. Lett.}
\newcommand{\apjs}{Astrophys. J. Suppl. Ser.}
\newcommand{\aap}{Astron. \& Astrophys.}
\newcommand{\aj}{Astron. J.}
\newcommand{\mnras}{Mon. Not. R. Astron. Soc.}
\newcommand{\physrep}{Phys. Rept.}

\newcommand{\lya}{Ly$\alpha$ }

      \newcommand{\ud}{\mbox{d}}
 
\begin{document}
 
   \title{Constraining the unexplored period between reionization and the dark ages with observations of the global 21 cm signal}
 
 \author{Jonathan R.~Pritchard}\thanks{Hubble Fellow}
 \email{jpritchard@cfa.harvard.edu}
\affiliation{Harvard-Smithsonian Center for Astrophysics, MS-51, 60 Garden St, Cambridge, MA 02138}
 \author{Abraham Loeb}
 \affiliation{Harvard-Smithsonian Center for Astrophysics, MS-51, 60 Garden St, Cambridge, MA 02138}
 
 \begin{abstract}
 Observations of the frequency dependence of the global brightness temperature of the redshifted 21 cm line of neutral hydrogen may be possible with single dipole experiments.  In this paper, we develop a Fisher matrix formalism for calculating the sensitivity of such instruments to the 21 cm signal from reionization and the dark ages.  We show that rapid reionization histories with duration $\Delta z\lesssim 2$ can be constrained, provided that local foregrounds can be well modelled by low order polynomials.  It is then shown that observations in the range $\nu=50-100{\,\rm MHz}$ can feasibly constrain the \lya and X-ray emissivity of the first stars forming at $z\sim15-25$, provided that systematic temperature residuals can be controlled to less than 1 mK.  Finally, we demonstrate the difficulty of detecting the 21 cm signal from the dark ages before star formation.
\end{abstract}
 
 
 
\maketitle

\section{Introduction} 
\label{sec:intro}
The transition of the Universe from the dark ages following hydrogen recombination through to the epoch of reionization remains one of the least constrained frontiers of modern cosmology.  Observing the sources responsible for heating and ionizing the intergalactic medium (IGM) at redshifts $z\gtrsim6$ pushes current observational techniques to the limit.  Plans are underway to construct low-frequency radio telescopes, such as LOFAR\footnote{http://www.lofar.org/}, MWA\footnote{http://www.MWAtelescope.org/}, PAPER\footnote{\citet{parsons2009}}, and SKA\footnote{http://www.skatelescope.org/}, to observe the red-shifted 21 cm line of neutral hydrogen.  These experiments aim to map the state of the intergalactic medium via tomographic observations of 3D fluctuations in the 21 cm brightness temperature.  A simpler and significantly lower cost alternative to this would be measurements of the global 21 cm signal integrated over the sky \cite{mmr1997,shaver1999,sethi2005}, which can be achieved by single dipole experiments like EDGES \cite{bowman2007edges} or CoRE \cite{chippendale2005}.  Although such experiments are today in their infancy, their potential is large.    In this paper, we explore the potential for these global sky experiments to measure the 21 cm signal and constrain the high redshift Universe.

We may draw a historical analogy with the Cosmic Background Explorer (COBE), whose FIRAS instrument measured the blackbody spectrum of the cosmic microwave background (CMB) \cite{mather1994} while the DMR instrument measured the level of temperature fluctuations \cite{smoot1992}.  The precise measurement of a $T_{\rm CMB}=2.726{\,\rm K}$ blackbody spectrum placed tight constraints on early energy injection, since no Compton-$y$ or $\mu$-distortion were seen, and provided important evidence confirming the big bang paradigm.  The detection of angular fluctuations paved the way for more sensitive experiments such as BOOMERANG \cite{lange2001} and WMAP \cite{spergel2003}, which provided precision measurements of the CMB acoustic peaks.  While, at the moment, attention is focussed on experiments designed to measure 21 cm fluctuations, it is important not to neglect the possibility of measuring the global signal.

The evolution of the 21 cm signal is driven primarily by the amount of neutral hydrogen and the coupling between the 21 cm spin temperature and the gas temperature.  It is able to act as a sensitive thermometer when the IGM gas temperature is less than the CMB temperature placing constraints on energy injection that leads to heating.  For example, the first black holes to form generate X-rays, which heat the gas.  More exotic processes such as annihilating dark matter might have also been important.  Additionally, energy injection in the form of \lya production modifies the strength of the coupling.  This provides a way of tracking star formation, which will be the dominant source of \lya photons.  As we show, the spectral structure of the 21 cm signal is much richer than that of a blackbody so that many things can be learnt about the early Universe. Given the uncertainties, we develop a model approach based upon those physical features most likely to be present.

The single most important factor determining the sensitivity of dipoles to astrophysics will be their ability to remove galactic foregrounds \cite[e.g.][]{oh2003,dimatteo2004}.  Exploitation of spectral smoothness to remove foregrounds by fitting low order polynomials is key to avoiding throwing the signal away with the foreground.  To quantify this, we develop a simple Fisher matrix formalism and validate it against more detailed numerical parameter fitting.  This provides us with a way of quantitatively addressing the ability of global 21 cm experiments to constrain reionization and the astrophysics of the first galaxies \cite{loeb2010book}.  Similar work on the subject \cite{sethi2005} ignored the influence of foregrounds limiting its utility considerably.

Much of the power of this technique stems from the limitations of other observational probes.  While next generation telescopes such as JWST\footnote{http://www.jwst.nasa.gov/}, GMT\footnote{http://www.gmto.org/}, EELT\footnote{http://www.eso.org/sci/facilities/eelt/} or TMT\footnote{http://www.tmt.org/} may provide a glimpse of the Universe at $z\gtrsim12$ they peer through a narrow field of view and are unlikely to touch upon redshifts $z\gtrsim20$.  As we will show, 21 cm global experiments could potentially provide crude constraints on even higher redshifts at a much lower cost.

The structure of this paper is as follows.
In \S\ref{sec:physics}, we begin by describing the basic physics that drives the evolution of the 21 cm global signature and drawing attention to the key observable features.  We follow this in \S\ref{sec:foreground} with a discussion of the foregrounds, which leads into our presenting a Fisher matrix formalism for predicting observational constraints in \S\ref{sec:fisher}.  In \S\ref{sec:reion} and \S\ref{sec:astro} we apply this formalism to the signal from reionization and the first stars, respectively.  After a brief discussion in \S\ref{sec:darkages} of the prospects for detecting the signal from the dark ages before star formation, we conclude in \S\ref{sec:conclude}.

Throughout this paper where cosmological parameters are required we use the standard set of values $\Omega_m=0.3$, $\Omega_\Lambda=0.7$, $\Omega_b=0.046$, $H=100h\,\rm{km\,s^{-1}\,Mpc^{-1}}$ (with $h=0.7$), $n_S=0.95$, and $\sigma_8=0.8$, consistent with the latest measurements \citep{komatsu2009}.
\section{Physics of the 21 cm global signal} 
\label{sec:physics}

The physics of the cosmological 21 cm signal has been described in detail by a number of authors \citep{fob,pritchard2008} and we focus here on those features relevant for the global signal.  It is important before we start to emphasise our uncertainty in the sources of radiation in the early Universe, so that we must of necessity extrapolate far beyond what we know to make predictions for what we may find.  Nonetheless the basic atomic physics is well understood and a plausible understanding of the likely history is possible.

The 21 cm line frequency $\nu_{\rm21\,cm}=1420{\,\rm MHz}$ redshifts for $z=6-27$ into the range 200-50 MHz.  The signal strength may be expressed as a differential brightness temperature relative to the CMB
  \begin{multline}\label{tb}
 T_b=27  x_{\rm{HI}}\left(\frac{T_S-T_\gamma}{T_S}\right)\left(\frac{1+z}{10}\right)^{1/2}\\ \times(1+\delta_b)\left[\frac{\partial_r v_r}{(1+z)H(z)}\right]^{-1}\,\rm{mK},
 \end{multline}
where $x_{\rm HI}$ is the hydrogen neutral fraction, $\delta_b$ is the overdensity in baryons, $T_S$ is the 21 cm spin temperature, $T_\gamma$ is the CMB temperature, $H(z)$ is the Hubble parameter, and the last term describes the effect of peculiar velocities with $\partial_r v_r$ the derivative of the velocities along the line of sight.  Throughout this paper, we will neglect fluctuations in the signal so that neither of the terms $\delta_b$ nor the peculiar velocities will be relevant.  Fluctuations in $x_H$ and $\delta_b$ will be relevant for the details of the signal, but are not required to get the broad features of the signal, on which we focus here.  

\begin{figure}[htbp]
\begin{center}
\includegraphics[scale=0.4]{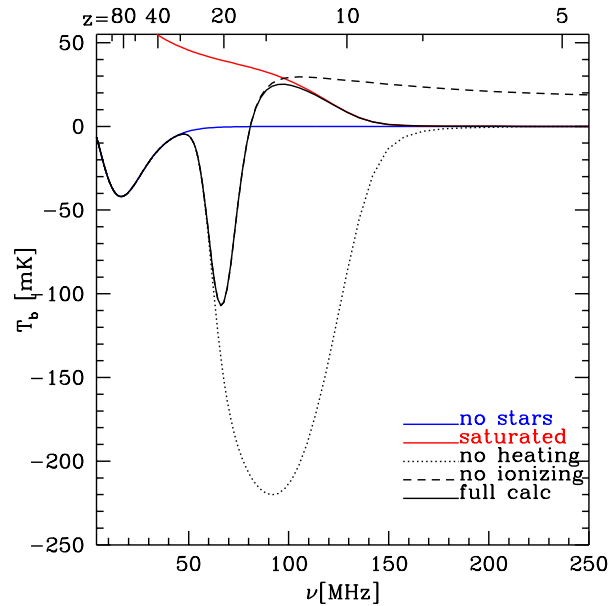}
\caption{Evolution of the 21 cm global signal for different scenarios.  {\em Solid blue curve: } no stars; {\em solid red curve: }$T_S\gg T_\gamma$; {\em black dotted curve: }no heating; {\em black dashed curve: }no ionization; {\em black solid curve: }full calculation.}
\label{fig:pedplot}
\end{center}
\end{figure}
The evolution of $T_b$ is thus driven by the evolution of $x_H$ and $T_S$ and is illustrated for redshifts $z<100$ in Figure \ref{fig:pedplot}.  Early on, collisions drive $T_S$ to the gas temperature $T_K$, which after thermal decoupling (at $z\approx1000$) has been cooling faster than the CMB leading to a 21 cm absorption feature ($[T_S-T_\gamma]<0$).  Collisions start to become ineffective at redshifts $z\sim80$ and scattering of CMB photons begins to drive $T_S\rightarrow T_\gamma$ causing the signal to disappear.  In the absence of star formation, this would be the whole story \cite{loeb_zald2004}.

Star formation leads to the production of \lya photons, which resonantly scatter off hydrogen coupling $T_S$ to $T_K$ via the Wouthysen-Field effect \cite{wouth1952,field1958}.  This produces a sharp absorption feature beginning at $z\sim30$.  If star formation also generates X-rays they will heat the gas, first causing a decrease in $T_b$ as the gas temperature is heated towards $T_\gamma$ and then leading to an emission signal, as the gas is heated to temperatures $T_K>T_\gamma$.  For $T_S\gg T_\gamma$ all dependence on the spin temperature drops out of equation \eqref{tb} and the signal becomes saturated.  This represents a hard upper limit on the signal.  Finally reionization will occur as UV photons produce bubbles of ionized hydrogen that percolate, removing the 21 cm signal.

\begin{figure}[htbp]
\begin{center}
\includegraphics[scale=0.4]{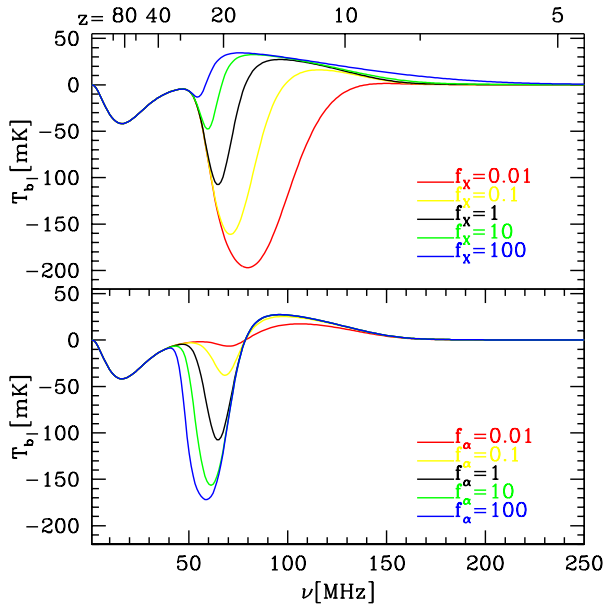}
\caption{Dependence of 21 cm signal on the X-ray (top panel) and \lya (bottom panel) emissivity.  In each case, we consider examples with the emissivity reduced or increased by a factor of up to 100.  Note that in our model $f_X$ and $f\alpha$ are really the product of the emissivity and the star formation efficiency.}
\label{fig:param_comp}
\end{center}
\end{figure}

We may thus identify five main events in the history of the 21 cm signal: (i) collisional coupling becoming ineffective (ii) \lya coupling becoming effective (iii) heating occurring (iv) reionization beginning (v) reionization ending.  In the scenario described above the first four of these events generates a turning point ($\ud T_b/\ud z=0$) and the final event marks the end of the signal.  We reiterate that the astrophysics of the sources driving these events is very uncertain, so that when or even if these events occur as described is currently  unknown.  Figure \ref{fig:param_comp} shows a set of histories for different values of the X-ray and \lya emissivity, parametrized about our fiducial model by $f_X$ and $f_\alpha$ representing the product of the emissivity and the star formation efficiency following Ref. \cite{pritchard2008}.  Clearly the positions of these features may move around both in the amplitude of $T_b$ and the frequency at which they occur. 

We view this to be the most likely sequence of events for plausible astrophysical models.  We are reassured in this sequencing since, in the absence of \lya photons escaping from galaxies \cite{higgins2009}, X-rays will also produce \lya photons \cite{chen2006,chuzhoy2006} and so couple $T_S$ to $T_K$ and, in the absence of X-rays, scattering of \lya photons heats the gas \cite{ciardi2003}.  In each case the relative sequence of events is likely to be maintained.  We will return to how different models may be distinguished later and now turn to the presence of foregrounds between us and the signal.

\section{Foregrounds} 
\label{sec:foreground}

\begin{figure}[htbp]
\begin{center}
\includegraphics[scale=0.25,angle=90]{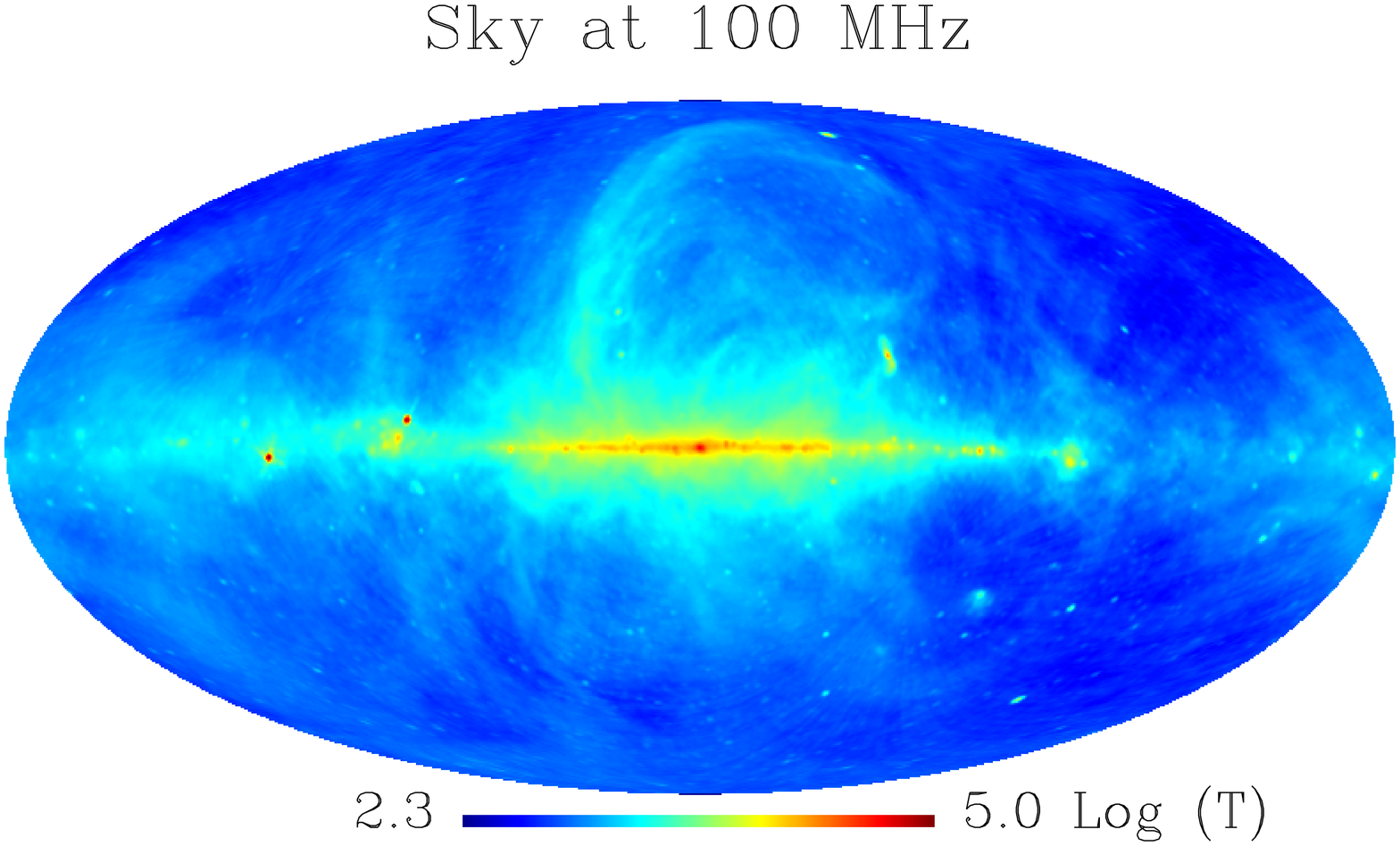}
\includegraphics[scale=0.25,angle=90]{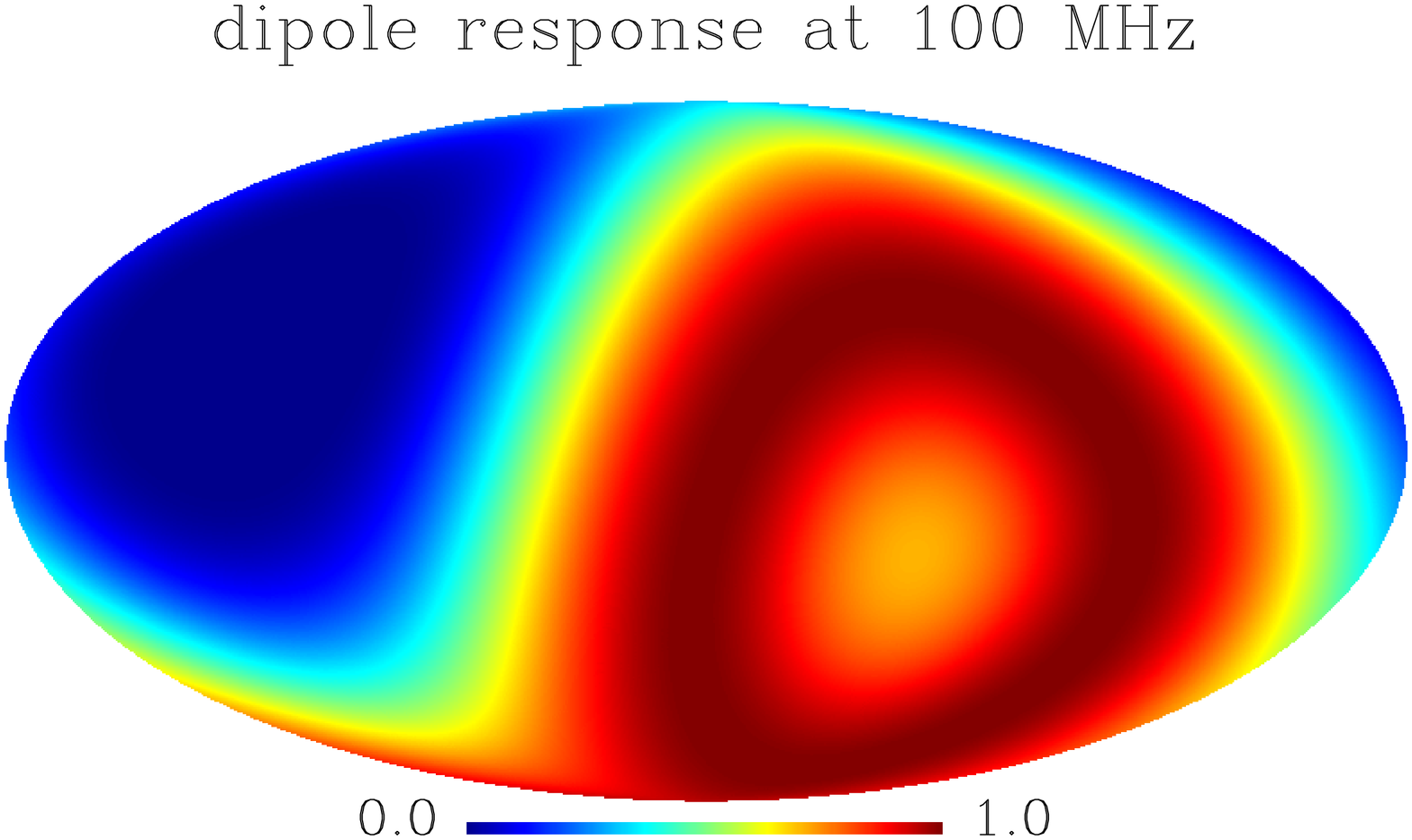}
\caption{{\em Top panel: }Radio map of the sky at 100 MHz generated from Ref. \cite{angelica2008}. {\em Bottom panel: } Ideal dipole response averaged over 24 hours.}
\label{fig:skymaps}
\end{center}
\end{figure}

At the frequencies of interest (10-250 MHz), the sky is dominated by synchrotron emission from the galaxy.  A useful model of the sky has been put together by Ref. \citep{angelica2008} using all existing observations.  The sky at 100 MHz is shown in Figure \ref{fig:skymaps}, where the form of the galaxy is clearly visible.  In this paper, we will be focusing upon observations by single dipole experiments.  These have beam shapes with a typical field-of-view of tens of degrees.  The lower panel of Figure \ref{fig:skymaps} shows the beam of dipole (approximated here as a single $\cos^2\theta$ lobe) sitting at the MWA site in Australia (approximate latitude 26$^\circ$59'S), observing at zenith, and integrated over a full day.  Although the dipole does not see the whole sky at once it does average over large patches.  We will therefore neglect spatial variations (although we will return to this point in our conclusions).

\begin{figure}[htbp]
\begin{center}
\includegraphics[scale=0.4]{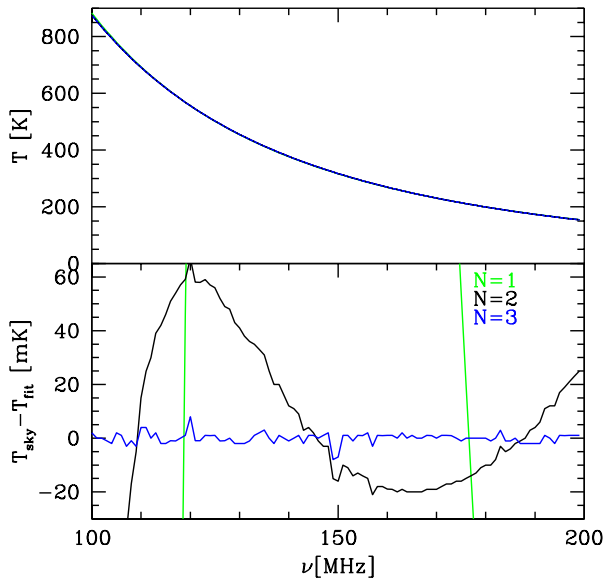}
\caption{Foreground (top panel) and residuals (bottom panel) left over after fitting a N-th order polynomial in $\log\nu$ to the foreground.}
\label{fig:fitresidual_nice}
\end{center}
\end{figure}

Averaging the foregrounds over the dipole's angular response gives the spectrum shown in the top panel of Figure \ref{fig:fitresidual_nice}.  First note that the amplitude of the foregrounds is large $\sim100{\rm\,K}$ compared to the 10 mK signal.  Nonetheless, given the smooth frequency dependence of the foregrounds we are motivated to try fitting the foreground out using a low order polynomial in the hope that this leaves the signal behind.  This has been shown by many authors \cite[e.g.][]{mcquinn2005,wang2006} to be a reasonable procedure in the case of 21 cm tomography.  There the inhomogeneities fluctuate rapidly with frequency, so that only the largest Fourier modes of the signal are removed.  In the case of the global 21 cm signal our signal is relatively smooth in frequency, especially if the bandwidth of the instrument is small.  Throwing the signal out with the foregrounds is therefore a definite concern.

Throughout this paper, we will fit the foregrounds using a polynomial of the form
\begin{equation}
\log T_{\rm fit}=\sum_{i=0}^{N_{\rm poly}}a_i \log(\nu/\nu_0)^i.
\end{equation}
Here $\nu_0$ is a pivot scale and we will generally recast $a_0\rightarrow \log T_0$ to emphasise that the zeroth order coefficient more naturally has units of temperature.  The lower panel of Figure \ref{fig:fitresidual_nice} shows the 
residuals left over after fitting and subtracting polynomials of different order to the foregrounds.  It is apparent that a polynomial of at least $N_{\rm poly}=3$ is necessary to remove the foreground.  Unfortunately, our current knowledge of the low frequency sky is not sufficient for us to conclusively say that we will not need a higher order polynomial or to accurately quantify the minimum level of residuals that will be left on fitting the signal.  The residuals visible in Figure \ref{fig:fitresidual_nice} for $N_{\rm poly}=3$ are dominated by numerical limitations of the sky model being used and have $\sqrt{\langle (T_{\rm sky}-T_{\rm fit})^2\rangle}\lesssim1{\,\rm mK}$ averaged over the band.


\begin{figure}[htbp]
\begin{center}
\includegraphics[scale=0.4]{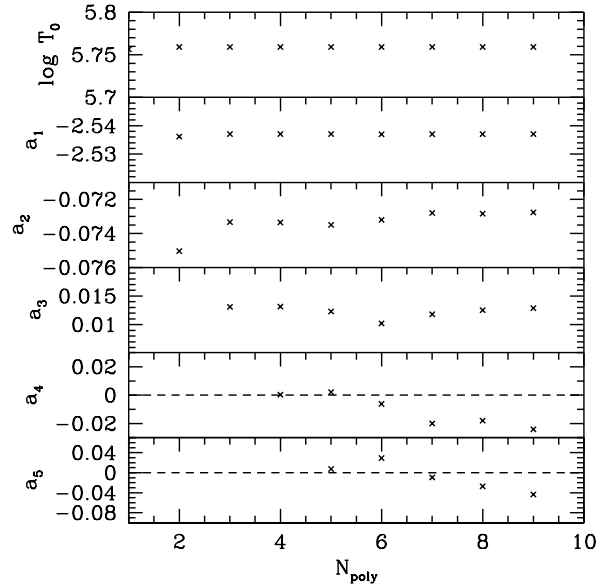}
\caption{Dependence of the best fit values for the first six parameters from the foreground fitting process on the order of the polynomial, $N_{\rm poly}$.}
\label{fig:fitvalues}
\end{center}
\end{figure}
Figure \ref{fig:fitvalues} shows the evolution of the best fit values as we change the order of the fit.  The first four values are non-zero and therefore important to the fit.  The next two hover around zero (although as the order increases they move away from zero).  This supports the inference that only the first four parameters are necessary and after that we are beginning to over fit.  We therefore take as our fiducial model for the foreground the form
\begin{multline}
\log T_{\rm sky}= \log T_0\\+a_1\log(\nu/\nu_0) + a_2[\log(\nu/\nu_0)]^2 +a_3 [\log(\nu/\nu_0)]^3,  
\end{multline}
with parameter values $\nu_0=150{\rm\,MHz}$, $T_0=320{\rm\,K}$, $a_1=-2.54$, $a_2=-0.074$, $a_3=0.013$, chosen from fitting to the band $\nu=100-200$ MHz.  These values are roughly consistent with those found by the observations reported in Ref. \cite{rogers2008}, which found $T_0=237\pm10{\,\rm K}$ and $a_1=-2.5\pm0.1$ over the same band.  Where necessary we include additional terms as $a_i=0$ for $i\geq4$.  Fitting to a different bandwidth and pivot frequency will modify these values.  For example, fitting to $\nu=50-150$ MHz with $\nu_0=100$ MHz yields, $T_0=875{\rm\,K}$, $a_1=-2.47$, $a_2=-0.089$, $a_3=0.013$.  Aside from the overall normalisation, there is little qualitative change in the shape.

\section{Fisher calculation} 
\label{sec:fisher}

The main objective of this paper is to develop a formalism for quantifying the ability of global 21 cm experiments to constrain astrophysical parameters.  A straightforward, but brute force approach, is to model the signal, add a foreground, and then use Monte-Carlo (MC) fitting techniques to see how well model parameters may be constrained.  When faced with the large space of model parameters to be explored this is inadequate.  We therefore explore the use of the Fisher matrix approach, applicable if the model likelihood is well approximated by a multivariate Gaussian.
We will later show that this is a good approximation by testing it directly against the results of direct MC fitting.

The Fisher matrix takes the form \cite{EHT99}
\begin{equation}
F_{ij}=\frac{1}{2}{\rm Tr}\left[C^{-1}C_{,i}C^{-1}C_{,j}+C^{-1}(\mu_{,i}\mu^T_{,j}+\mu_{,j}\mu^T_{,i})\right].
\end{equation}
where $C\equiv\langle x x^T\rangle$ is the covariance matrix and $\mu=\langle x\rangle$.  For the 21 cm global signature, our observable is the antennae temperature $T_{\rm sky}(\nu)=T_{\rm fg}(\nu)+T_{b}(\nu)$, where we assume the dipole sees the full sky so that spatial variation can be ignored.  We divide the signal into $N_{\rm channel}$ frequency bins \{$\nu_n$\} of bandwidth $B$ running between [$\nu_{\rm min}$, $\nu_{\rm max}$].  The covariance matrix is taken to be diagonal, since errors in different frequency bins are expected to be uncorrelated, so that it is given by
\begin{equation}
C_{ij}=\delta_{ij}\sigma_i^2,
\end{equation}
with the thermal noise given by the radiometer equation 
\begin{equation}
\sigma_i^2=\frac{T_{\rm sky}^2(\nu_i)}{B t_{\rm int}},
\end{equation}
assuming an integration time $t_{\rm int}$.  In this paper, we will consider single dipole experiments, but the noise could be further reduced by a factor $N_{\rm dipole}$ through the incoherent summing of the signal from multiple dipoles.  Finally, we can allow for a limiting floor in the noise due to foreground fitting residuals or instrumental noise by setting $\sigma_i^2\rightarrow\sigma_i^2+\sigma_{i,{\rm res}}^2$.

Under these assumptions the Fisher matrix takes the form
\begin{equation}
F_{ij}=\sum_{n=1}^{N_{\rm channel}}(2+B t_{\rm int})\frac{\ud\log T_{\rm sky}(\nu_n)}{\ud p_i}\frac{\ud\log T_{\rm sky}(\nu_n)}{\ud p_j},
\end{equation}
where the parameter set $\{p_i\}$ includes both foreground and signal model parameters.  Here the first term is the information contained in the amplitude of the noise and is subdominant for reasonable experiments (\cite[cf.][]{sethi2005}).  Given this Fisher matrix, the best parameter constraints achievable on parameter $p_i$ are given by the Cramer-Rao inequality $\sigma_i\geq\sqrt{F^{-1}_{ii}}$.
This Fisher matrix offers a fast and, as we will show in the next section, reliable means of calculating the expected constraints for 21 cm global experiments.

The assumption of a full sky observation is not strictly valid, since the dipole sees the sky with a beam tens of degrees across.  Both foreground and signal will show spatial variation.  Fluctuations in the 21 cm signal can be large in amplitude, but span a characteristic scale of order a few arcminutes corresponding to the size of the ionized bubbles.  As such our beam will average over many of these, so that we do not expect significant spatial fluctuations to survive.  The foregrounds are another matter and spatial variation may be a mixed blessing.  In practice, each foreground parameters should be fitted independently in each pixel.  Since the signal is common to all pixels, exploiting the spatial variation of the foregrounds could be used to remove them more efficiently.

So far, we have assumed that the instrument's frequency response can be calibrated out perfectly.  At present one of the limiting factors of the EDGES experiment is that the dipole's frequency response is uncalibrated.  This has the effect of convolving both foregrounds and signal with some unknown function of frequency.  Provided that this function is smooth the main complication so introduced is that the convolved foregrounds are no longer easily described by a low order polynomial.  In Ref. \cite{bowman2007edges}, a 12th order polynomial in $\nu$ was used for the foreground fitting, primarily in order to fit out the instrumental response.  Since this is very much a prototype experiment, we will optimistically assume that this instrumental problem can be dealt with in more advanced designs.


\section{Reionization} 
\label{sec:reion}

Next, we will consider the possibility of constraining the evolution of the hydrogen neutral fraction from the global 21 cm signal.  Predicting the reionization history has attracted a great deal of attention in recent years \cite{loeb2010book}.  Constraints arise from the \lya forest, the optical depth to the CMB, and numerous other locations.  Although these may be combined to constrain the reionization history \cite[e.g.][]{pritchard2009}, the quality of current constraints is poor.  In general though, reionization is expected to be a relatively extended process.

\begin{figure}[htbp]
\begin{center}
\includegraphics[scale=0.4]{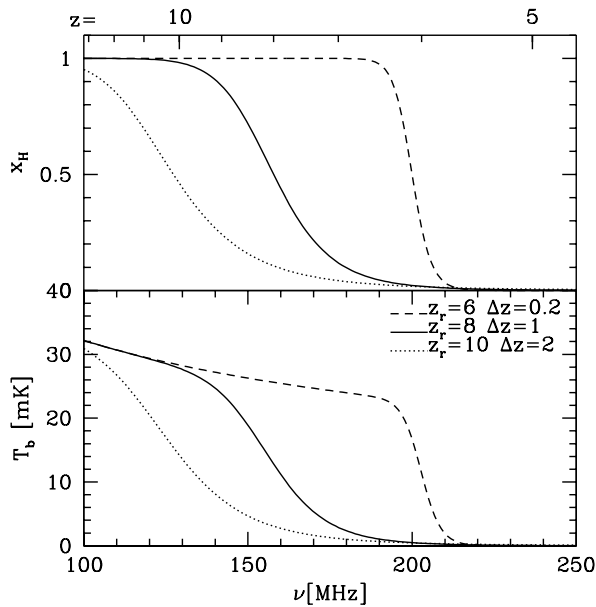}
\caption{Evolution of the neutral fraction $x_H$ and brightness temperature $T_b$ for a {\em tanh} model of reionization (see Eq.\ref{tanh}).}
\label{fig:nuhistory_s}
\end{center}
\end{figure}

Given the uncertainty associated with making detailed predictions for the evolution of $x_H$, we adopt as a toy model for reionization a {\em tanh} step (as used by the WMAP7 analysis \cite{larson2010}) with parameters describing the two main features of reionization: its mid point $z_r$ and duration $\Delta z$.  We will further assume that the 21 cm spin temperature is saturated at the relevant redshifts (a reasonable although not guaranteed simplifying assumption \cite{ciardi2003,pritchard2007xray}). Under these assumptions, the 21 cm brightness temperature is given by
\begin{equation}\label{tanh}
T_{b}(z)=\frac{T_{21}}{2}\left(\frac{1+z}{10}\right)^{1/2}\left[\tanh\left(\frac{z-z_r}{\Delta z}\right)+1\right].
\end{equation}
In principle, the amplitude of the signal $T_{21}$ is calculable from first principles ($T_{21}=27{\,\rm mK}$ for our fiducial cosmology), but we leave it as a free parameter.  This helps us gauge how well the experiment is really detecting the 21 cm signal.  Figure \ref{fig:nuhistory_s} shows a few different histories for this model.

\begin{figure}[htbp]
\begin{center}
\includegraphics[scale=0.5]{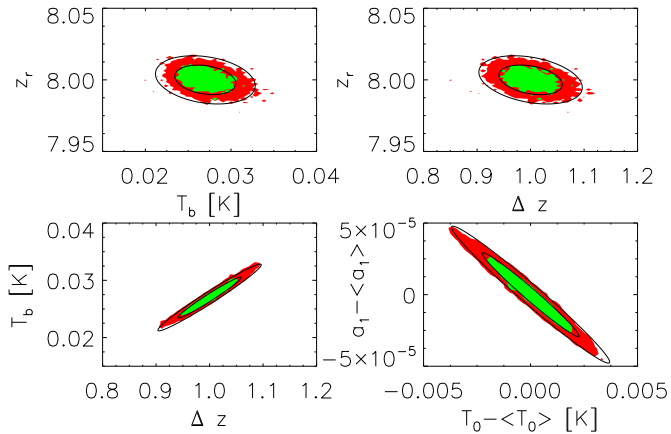}
\caption{Comparison of 68 and 95\% confidence regions between our MC likelihood (green and red coloured regions) and Fisher matrix (solid ellipses) calculations for a {\em tanh} model of reionization with $z_r=8$ and $\Delta z=1$ and fitting four foreground parameters.}
\label{fig:mc_comp}
\end{center}
\end{figure}

Before exploring the detection space for 21 cm experiments, we validate our Fisher matrix against a more numerically intensive Monte-Carlo.  We consider an experiment covering the frequency range $100-250{\,\rm MHz}$ in 50 bins and integrating for 500 hours (these parameters mimic EDGES with an order of magnitude longer integration time). Taking fiducial values of $z_r=8$, $\Delta z=1$, and $N_{\rm poly}=3$, we fit the model and foreground for $10^6$ realisations of the thermal noise.  This yields an estimate of the parameter uncertainty that can be expected from observations and can be used to test our Fisher matrix calculation.  The resulting parameter contours are shown in Figure \ref{fig:mc_comp} along with the Fisher matrix constraints.  That they are in good agreement validates our underlying formalism.  

The error ellipses show that there is a strong degeneracy between $T_{21}$ and $\Delta z$.  This is a consequence of the way in which foreground fitting removes power from more extended histories making it difficult to distinguish a larger amplitude extended scenario from a lower amplitude sharper scenario.

Despite the good agreement, this formalism breaks down when the Fisher matrix errors become large enough that reionization parameters are not well constrained. Although this is not a major hurdle here, caution should be used when errors are much larger than the parameters being constrained.


\begin{figure}[htbp]
\begin{center}
\includegraphics[scale=0.5]{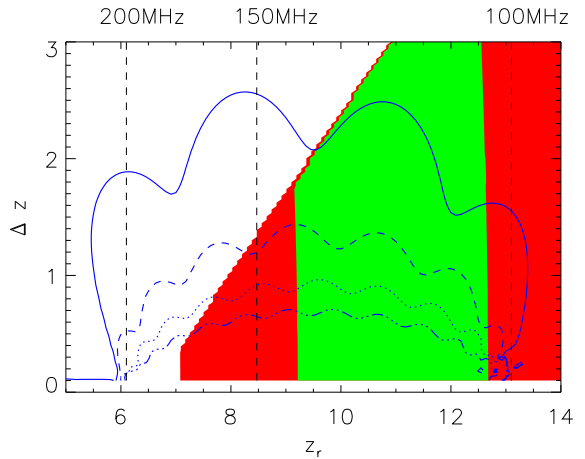}
\caption{95\% detection region for global experiments assuming $N_{\rm poly}=3$ (solid curve), 6 (dashed curve), 9 (dotted curve), and 12 (dot-dashed curve). Also plotted are the 68 and 95\% contours for WMAP5 with a prior that $x_i(z=6.5)>0.95$ (green and red coloured regions).}
\label{fig:zr_dz}
\end{center}
\end{figure}
The resulting potential detection region for the above experiment is shown in Figure \ref{fig:zr_dz}, where we consider several different orders of polynomial fit.  The detection region shows a number of wiggles associated with points in the frequency range where the shape of the 21 cm signal becomes more or less degenerate with the polynomial fitting.  We also show the 1- and $2-\sigma$ constraint regions from WMAP's optical depth measurement.  These constrain the redshift of reionization, but say little about how long it takes.  Adding in a prior based upon \lya forest observations that the Universe is fully ionized by $z=6.5$ (specified here as $x_i(z=6.5)>0.95$) removes the region of parameter space with large $\Delta z$ and low $z_r$.

Global experiments can take a good sized bite out of the remaining parameter space.  They are sensitive to the full range of redshifts, but primarily to the sharpest reionization histories.  Only if $N_{\rm poly}\le6$ can histories with $\Delta z>1$ be constrained and histories with $\Delta z\gtrsim2.5$ appear too extended for high significance detections.

This is unfortunate, since \citet{pritchard2009} found that most reionization histories compatible with the existing data have $\Delta z\gtrsim2$, suggesting it will be difficult for global experiments to probe the most likely models.  
An important caveat to these conclusions is that the {\em tanh} model that we have used here is a toy model of reionization.  More realistic models may have more detectable features since they often end rapidly, but have a long tail to high redshifts.

\section{First sources} 
\label{sec:astro}

We now turn from reionization to the signal produced by the first galaxies, which generate an early background of \lya and X-ray photons.  This region is essentially unconstrained by existing observations and global 21 cm experiments represent one of the only upcoming ways of probing this epoch.

\begin{figure}[htbp]
\begin{center}
\includegraphics[scale=0.4]{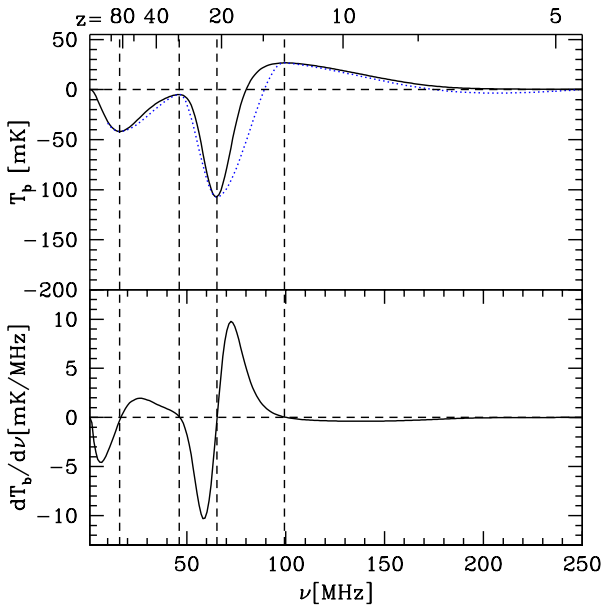}
\caption{Evolution of the 21 cm global signal and its derivative.  Vertical dashed lines indicate the locations of the turning points.  In the top panel, we also show a cubic spline fit to the turning points (blue dotted curve) as described in the text.}
\label{fig:spline_comp}
\end{center}
\end{figure}

Although models for the signal during this epoch exist \citep{furlanetto2006,pritchard2008}, it will be useful to focus on physical features of the signal that are both observable and model independent.  With this in mind, we parametrize the signal in terms of the turning points of the 21 cm signal.  Figure \ref{fig:spline_comp} shows the evolution of $T_b$ and its frequency derivative.  As discussed in \S\ref{sec:physics}, there are four turning points associated with: (0) a minimum during the dark ages where collisional coupling begins to become ineffective, (1) a maximum at the transition from the dark ages to the \lya pumping regime as \lya pumping begins to be effective, (2) an absorption minimum as X-ray heating begins to raise the signal towards emission, (3) an emission maximum as the signal becomes saturated and starts to decrease with the cosmic expansion.  Finally reionization completes providing a fifth point.  Asymptotically the signal goes to zero at very low and high frequencies.

In order to have a simple model for the evolution of the signal, we adopt parameters $(\nu_0,T_{b0})$, $(\nu_1,T_{b1})$, $(\nu_2,T_{b2})$, $(\nu_3,T_{b3})$, and $\nu_4$ for the frequency and amplitude of the turning points and the frequency at the end of reionization.  For clarity of notation we will label these points as $\mathbf{x}_i=(\nu_i,T_{bi})$ (with $\mathbf{x}_4=(\nu_4,0{\,\rm mK})$).  We then model the signal with a simple cubic spline between these points with the additional condition that the derivative should be zero at the turning points (enforced by doubling the data points at the turning points and offsetting them by $\Delta\nu=\pm1{\,\rm MHz}$).  

For our fiducial model, we adopt the fiducial parameter set of Ref. \citep{pritchard2008}, assuming a star forming efficiency $f_*=0.1$, a \lya emissivity expected for Population II stars $f_\alpha=1$, and X-ray emissivity appropriate for extrapolating the locally observed X-ray-FIR correlation, $f_X=1$.  This gives turning points $\mathbf{x}_0$=(16.1 MHz, -42 mK), $\mathbf{x}_1$=(46.2 MHz, -5 mK), $\mathbf{x}_2$=(65.3 MHz, -107 mK), $\mathbf{x}_3$=(99.4 MHz, 27 mK), and $\mathbf{x}_4$=(180 MHz, 0 mK).  The resulting spline fit is shown in the top panel of Figure \ref{fig:spline_comp}.  The model does a good job of capturing the general features of the 21 cm signal, although there are clear differences in the detailed shape.  Since global experiments are unlikely to constrain more than the sharpest features, this approach should be adequate for our purposes. 

\begin{figure}[htbp]
\begin{center}
\includegraphics[scale=0.5]{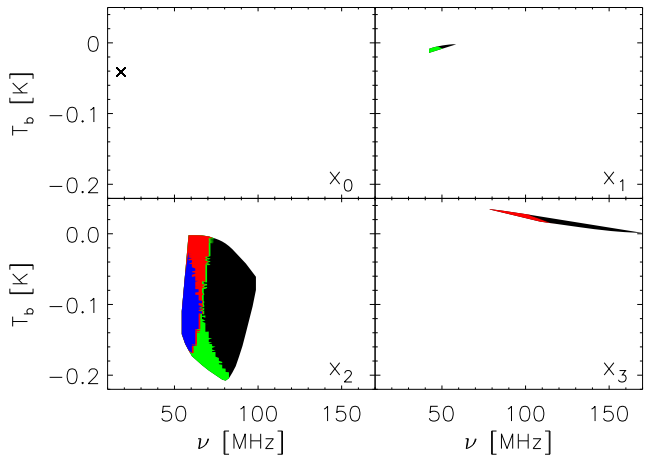}
\caption{Parameter space for the frequency and brightness temperature of the four turning points of the 21 cm signal calculated by varying parameters over the range $f_X=[0.01,100]$ and $f_\alpha=[0.01,100]$ for fixed cosmology and star formation rate $f_*=0.1$. Green region indicates $f_\alpha>1$, red region indicates $f_X>1$, blue regions indicates both $f_\alpha>1$ and $f_X>1$, while the black region has $f_\alpha<1$ and $f_X<1$.}
\label{fig:turning_map}
\end{center}
\end{figure}

There is considerable uncertainty in the parameters of this model, and so to gauge the likely model dependence of the turning points, we make use of the model of Ref. \citep{pritchard2008}.  Varying the \lya, X-ray, and UV emissivity by two orders of magnitude on either side of their fiducial values we find the position and amplitude of the turning points to give the parameter space shown in Figure \ref{fig:turning_map}.  This provides a useful guide to targeting observations in frequency space.  We have found that a global experiment has very little sensitivity to features lying outside of the observed frequency band.

Since we fix the cosmology, $\mathbf{x}_0$ appears as a single point.  The locations of $\mathbf{x}_1$ and $\mathbf{x}_3$ are controlled by the \lya and X-ray emissivity respectively.  Only $\mathbf{x}_2$ shows significant dependence on both \lya and X-ray emissivity leading to a large uncertainty in its position.  This is good news observationally, since even a poor measurement of the position of $\mathbf{x}_2$ is likely to rule out a wide region of parameter space.  Since $\mathbf{x}_2$ is the feature with both the largest amplitude and sharpest shape, we expect that this is the best target for observation and makes experiments covering $\nu=50-100{\,\rm MHz}$ of great interest.

\begin{figure}[htbp]
\begin{center}
\includegraphics[scale=0.5]{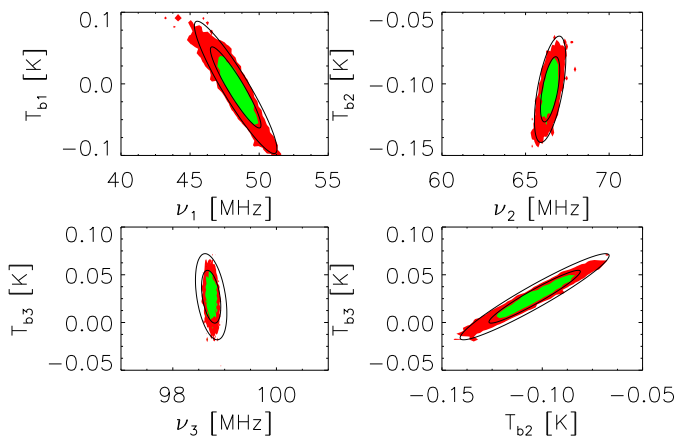}
\caption{Comparison of the 68 and 95\% confidence regions for our MC likelihood (green and red coloured regions) and Fisher matrix (solid contours).  The MC calculation fits the turning point model to the full signal while the Fisher matrix calculation is for the turning point model only.}
\label{fig:mc_comp_high_full}
\end{center}
\end{figure}

Since our model is approximate, it is important to check whether it leads to significantly biased constraints on the features of interest.  One could imagine that fitting the splined shape might lead to biased estimates of the position of the turning points, for example.  We have checked this through Monte-Carlo simulation by fitting the turning-point model to the full calculation signal for $10^6$ realisations of the thermal noise.  As seen in Figure \ref{fig:mc_comp_high_full} for an experiment covering $\nu=$45-145 MHz in 50 bins and integrating for 500 hours, the MC calculation shows no sign of significant biasing and is in good agreement with the Fisher matrix calculation using the turning-point model.

The final panel of Figure \ref{fig:mc_comp_high_full} shows a degeneracy between $T_{b2}$ and $T_{b3}$.  This might be expected for an experiment whose sensitivity is primarily to the derivative of the signal, which is left unchanged by shifting both of these points up or down.

\begin{figure}[htbp]
\begin{center}
\includegraphics[scale=0.4]{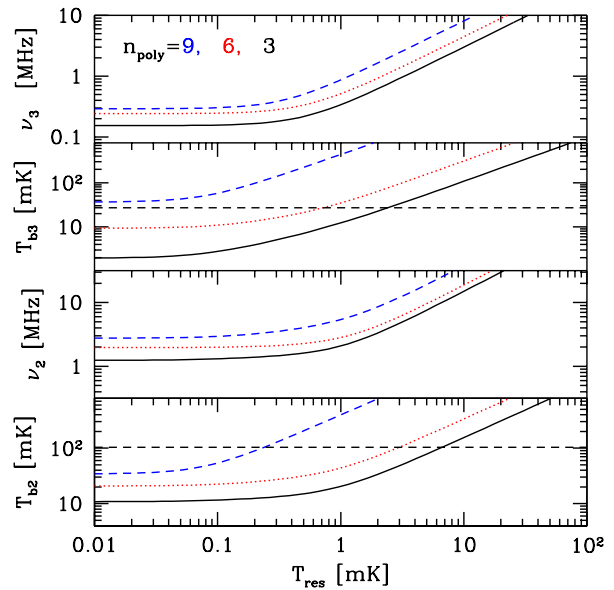}
\caption{Dependence of ($\nu_3,T_{b3}$) and ($\nu_2,T_{b2}$) errors with level of systematic residuals for $N_{\rm poly}=3$ (black solid curve), 6 (red dotted curve), and 9 (blue dashed curve).  The dashed vertical lines indicates the fiducial values $T_{b3}=27{\,\rm mK}$ and $|T_{b3}|=107{\,\rm mK}$.}
\label{fig:fgerror_tsys_nu3}
\end{center}
\end{figure}

As we examine lower frequencies where the foregrounds are brighter, we must increasingly worry about foreground removal leaving behind systematic residuals that limit the sensitivity of the experiment.  In Figure \ref{fig:fgerror_tsys_nu3}, we plot the sensitivity of the same experiment to $\mathbf{x}_3$ as a function of this residual floor $T_{\rm res}$ for different values of $N_{\rm poly}$.  Polynomials with $N_{\rm poly}<9$ are required to have any chance of detecting the signal.  Sensitivity to the signal begins to degrade once $T_{\rm res}$ becomes greater than 0.1 mK corresponding roughly to the thermal noise for this experiment.  A detection of $\mathbf{x}_3$ is still possible until $T_{\rm sys}\sim1{\,\rm mK}$.

\begin{figure}[htbp]
\begin{center}
\includegraphics[scale=0.5]{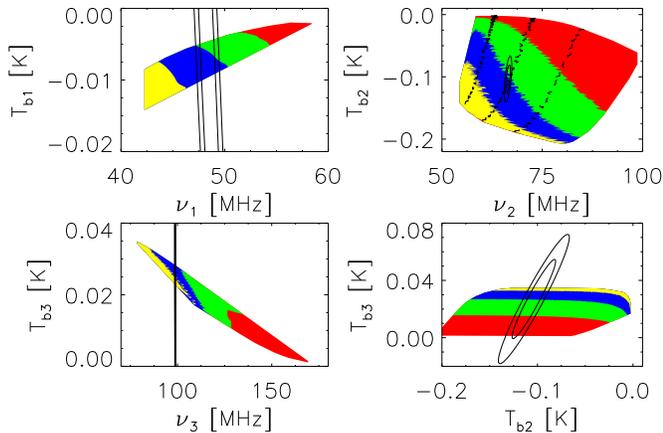}
\caption{Experimental constraints overlaid on the allowed region for the turning points.  Shaded regions (dashed curves) illustrate contours of $f_X$ and $f_\alpha$ by an order of magnitude (red to yellow).}
\label{fig:turning_contour}
\end{center}
\end{figure}

We finish this section by comparing the Fisher matrix constraints from Figure \ref{fig:mc_comp_high_full} on top of the region spanned by the turning points in Figure \ref{fig:turning_map}.  This is shown in Figure \ref{fig:turning_contour} and gives a sense of the large space of astrophysical models that may be ruled out with a single global experiment.  While the experiment has trouble constraining $\mathbf{x}_1$ and $\mathbf{x}_3$ with any significance, it places relatively good constraints on $\mathbf{x}_2$.  

Throughout this section we have chosen to model the 21 cm global signal by a simple cubic spline based upon the turning points of the signal.  While this model is simple, one can imagine alternative approaches.  Since the experiments are primarily sensitive to the derivative of the 21 cm signal, we might imagine taking the positions of the extrema of the derivative $\ud T_b/\ud \nu$ as our parameters and seek to constrain those.  We leave the exploration of alternatives such as this to future work.

\section{Dark Ages} 
\label{sec:darkages}

The physics of the period before star formation at $z\sim30$ is determined by well known atomic processes and so has much in common with the CMB.  However, many models have been put forward that would modify this simple picture with exotic energy deposition via annihilating or decaying  dark matter \cite{furlanetto2006dm} or evaporating black holes \cite{mack2008}, for example.  During the dark ages, the 21 cm signal acts as a sensitive thermometer, potentially capable of constraining these exotic processes.  Here we will focus on the standard history and leave consideration of the possibility of detecting other scenarios to future work.

The signal during the dark ages reaches a maximum at $\mathbf{x}_0=(16{\,\rm MHz}$, $-42{\,\rm mK})$, somewhat larger in amplitude than the reionization emission signal.  However, at these low frequencies the foregrounds are extremely large, $T_{\rm fg}\approx10^4{\,\rm K}$ at $\nu=30{\,\rm MHz}$, making detection very difficult. Its is worth noting however that global experiments have an advantage over tomographic measurements here, since at these early times structures have had little chance to grow, making the fluctuations much smaller than during reionization.  Further, it is easier to imagine launching a single dipole experiment beyond the Earth's ionosphere rather than the many km$^2$ of collecting area needed for interferometers to probe this epoch \cite{gordon2009,jester2009}.

Given the large foregrounds, long integration times or many dipoles are required to reach the desired sensitivity level.  Taking $T_{\rm fg}=10^4{\,\rm K}$ at $\nu=30{\,\rm MHz}$ a single dipole would need to integrate for $t_{\rm int}=1000{\,\rm hours}$ to reach 4 mK sensitivity.  Removing the foregrounds over this dynamic range without leaving considerable residuals will clearly require very precise instrumental calibration.  Given the challenges, we look at the most optimistic case as a limit of what could be accomplished.

\begin{figure}[htbp]
\begin{center}
\includegraphics[scale=0.5]{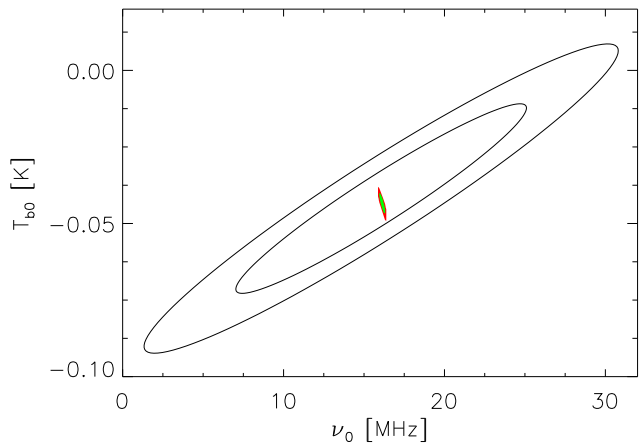}
\caption{68 and 95\% error ellipses on the amplitude and frequency of the dark ages minima for a single dipole experiment (solid curves, see text for details).  For comparison, we show the spread in these quantities from the WMAP5 1- and 2-$\sigma$ uncertainties in $\Omega_m h^2$ and $\Omega_b h^2$ (green and red coloured region).}
\label{fig:mc_comp_dark}
\end{center}
\end{figure}

Taking an experiment covering $\nu=5-60{\,\rm MHz}$ in 50 channels and integrating for 8000 hours, we assume a minimal $N_{\rm poly}=3$ polynomial fit leaving no residuals.  The resulting constraint on the position and amplitude of the dark ages feature are shown in Figure \ref{fig:mc_comp_dark}.  Such an experiment is capable of detecting the signal, but only barely.  For comparison, we have plotted the uncertainty arising from cosmological measurements of $\Omega_mh^2$ and $\Omega_bh^2$, the two main parameters determining the 21 cm signal.  This uncertainty is much less than the experimental uncertainty.

Although we have shown that detecting the dark ages feature from the standard history would be extremely challenging, modified histories arising from exotic energy injection may lead to larger features more easily detected.  Since there is no other probe of physics at $30<z<150$ global 21 cm experiments offer a unique if extremely challenging probe of this period. 

\section{Conclusions} 
\label{sec:conclude}

Observations of the redshifted 21 cm line potentially provide a new window into the high redshift Universe.  Detecting this signal in the presence of large foregrounds is challenging and it is important to explore all avenues for exploiting the signal.  In this paper, we have focussed upon the possibility of using single dipole experiments to observe the all-sky 21 cm signal, in contrast to the 21 cm fluctuations targeted by MWA, LOFAR, PAPER, and SKA.  Experiments targeting this global signal are in their infancy.  We emphasise that instruments built from a few dipoles targeting the global 21 cm signal can be several orders of magnitude cheaper to build than interferometers targeting the fluctuations.  Their scientific return will be similarly less, but at this stage where we know so little about the first sources, even that little is extremely valuable.

As we have outlined in this paper, the 21 cm signal generated by astrophysical processes has a well defined form, although the input parameters are only poorly understood.  We have demonstrated that, at the level of our current knowledge, describing the Galactic foregrounds requires at least a 3rd order polynomial.  At this level, we are able to remove the foregrounds to the sub-mK level, although in practice this procedure may be more complicated.  In order to characterise the sensitivity of these experiments to the signal, we developed a Fisher matrix formalism and validated it against more numerical fitting of the model parameters.  This Fisher matrix approach allows rapid calculations of the experimental sensitivity and appears to reproduce more detailed calculations very well.  

Having developed this formalism we applied it to the signal from reionization and the epoch of the first stars.  Using a toy model of reionization, we demonstrated that EDGES-like experiments should be capable of constraining rapid reionization histories with $\Delta z\lesssim2$.  More promisingly, these experiments can rule out a wide variety of astrophysical models for the signal from the first stars where the evolution of the spin temperature is important.  We used a straightforward fitting form for the signal based upon the positions of the turning points and showed that these features could be constrained, with the deepest absorption trough providing the best observational target.

Finally, we briefly explored the possibility of detecting the absorption feature present before star formation began.  The increased foreground brightness at low frequencies make it very difficult to constrain this feature and will require long integration times and more sophisticated methods of foreground removal.

This paper represents a first serious look at the prospects for using global measurements of the 21 cm signal to constrain astrophysics.  As a result, there are a number of places where future work might improve upon our calculations.  These include investigating the effects of finite sky coverage, incorporating an arbitrary instrumental frequency response, and allowing for the removal of frequency channels corrupted by terrestrial radio interference.

\section*{Acknowledgements}

We would like to thank Judd Bowman and Angelica de Oliveira-Costa for useful conversations.  Figure \ref{fig:skymaps} was generated using HEALpix \cite{gorski2005} and the global sky model software of Ref. \cite{angelica2008}.
JRP is supported by NASA through Hubble Fellowship grant HST-HF-01211.01-A
awarded by the Space Telescope Science Institute, which is operated by the
Association of Universities for Research in Astronomy, Inc., for NASA,
under contract NAS 5-26555. AL acknowledges funding from NSF grant AST-0907890 and NASA grants NNA09DB30A and NNX08AL43G.


 
 \end{document}